\documentclass[twocolumn,showpacs,preprintnumbers,amsmath,amssymb]{revtex4}
\usepackage{graphicx}
\usepackage{dcolumn}
\usepackage{bm}

\begin{document}

\title{Electronic structure of HgBa$_2$CuO$_{4+\delta}$ with self-organized interstitial oxygen wires in the
Hg spacer planes}

\author{Thomas Jarlborg$^{1,2}$ and Antonio Bianconi$^{2,3,4}$}

\affiliation{
$^1$DPMC, University of Geneva, 24 Quai Ernest-Ansermet, CH-1211 Geneva 4,
Switzerland
\\
$^2$ RICMASS Rome International Center for Materials Science Superstripes, Via dei Sabelli 119A, 00185 Rome, Italy
\\
$^3$ Institute of Crystallography, Consiglio Nazionale delle Ricerche, CNR, Via Salaria Km 29.300, Monterotondo, Roma, I-00015, Italy
\\
$^4$ National Research Nuclear University MEPhI (Moscow Engineering Physics Institute),  Kashirskoe sh. 31, 115409 Moscow, Russia }


\begin{abstract}

While recent experiments have found that at optimum doping for the highest critical temperature 
in HgBa$_2$CuO$_{4+y}$ (Hg1201) the oxygen interstitials (O-i)  
are not homogeneously distributed but form one-dimensional atomic wires,
there are no available information of its electronic structure  considering self-organized O-i atomic wires.
Here we report the calculated electronic structure of HgBa$_2$CuO$_{4+y}$  where oxygen interstitials form atomic wires 
along (1,0,0) crystal direction in the Hg layer.  We find that at optimum doping for superconductivity the chemical potential 
is tuned near an electronic topological Lifshitz transition for the appearing of a second quasi 1D Fermi surface. 
A $first$ large Fermi surface coexists with a $second$ incipient quasi one dimensional (1D) 
Fermi surface related with atomic wires of oxygen interstitials.  
Increasing oxygen doping the chemical potential is driven to the band edge of the $second$ 
1D-band giving a peak in the density-of-states. The new 1D electronic states are confined near the oxygen interstitial wires 
 with a small spread only on nearby sites. Spin-polarized calculations show that the magnetic response 
 is confined in the oxygen-poor domains free of oxygen interstitials wires and it is quite insensitive to the density of O-i wires. 

\end{abstract}


\maketitle

\section{Introduction.}

In these last years novel experiments have observed that the self-organization of interstitials and defects
can dramatically enhance or suppress the critical temperature $T_c$ in cuprates \cite{little,frat,poccia1}.
The self-organization of defects and interstitials in the spacer layers depends
 on the lattice misfit $strain$ between the spacer layers and the Cu$O_2$ layers.
The lattice strain has been proposed to be the third axis of the 3D phase diagram
 $T_c$, $doping$. $strain$  for all cuprates families \cite{dicastro}. 
 In  $Bi_2$$Sr_2$$Ca$$Cu_2$$O_{8+y}$ (Bi2212) the misfit strain induces a relevant 
 corrugation of the Cu$O_2$ planes  observed by Cu K-edge Resonant Elastic X-ray Scattering (REXS) \cite{lus} 
 accompanied by the superconducting gap modulation  \cite{sle,and}.
The formation of stripes in cuprates has been first observed using synchrotron radiation x-ray spectroscopy 
which probes the local structure in
 $Bi_2$$Sr_2$$Ca_{1-x}$$Y_x$$Cu_2$$O_{8+y}$  \cite{pom,miss,bia1,bia2} 
 due to cooperative effects of misfit strain
and CDW formation with wave-vector depending on doping \cite{agrestini,kus}.
The investigation of the variation of the critical temperature in complex defective 
transition metal oxides as a function of interstitials organization
has provided  evidence for defects and interstitials organization
in $Na_{x}$Co$O_{2}$+y$D_2$O  (x=1/3; y=4x) \cite{barn} and in 
$Sr_3$$Co_2$$O_x$ ($5.64<x<6.60$) \cite{hill}.
The defects self organization controlling the critical temperature has been found in $Sr_2$$CuO_{4-y}$  \cite{geba}, in
$Sr_{2-x}$$Ba_x$Cu$O_{3+y}$  \cite{gao},   in ($Cu_{0.75}$$Mo_{0.25}$$Sr_2$Y$Cu_2$$O_{7+y}$  
with $0<y<0.5$  \cite{chma1,chma2},  and in $BaPb_{1-x}Bi_xO_3$ \cite{gallo}.
The oxygen interstitial organization in  oxygen doped cuprates $La_2$Cu$O_{4+y}$ 
has been studied by scanning micro x-ray diffraction    
 \cite{frat,poccia1,ricci1,ricci2,ricci3,ricci4,campi1,campi2},  and by STM  \cite{zel1,zel2,zel3}
showing superconductivity emerging in a nanoscale phase separation with a
complex geometry \cite{superstripes,kresin,gin12,gin12a,gin13}. 
The phase separation is determined by the proximity to a electronic topological Lifshitz transition 
in strongly correlated electronic systems \cite{kugel1,kugel2,kugel3} .
There is now growing agreement that the domes of high critical temperature in different superconductors occur 
where the chemical potential is tuned near topological electronic Lifshitz transitions 
 \cite{lifshitz1,lif-cup-1,lif-cup-2,lif_a,lif_b}
 including the case of pressurized sulfur hydride  \cite{lif-h3s-1,lif-h3s-2,lif-h3s-3}.
 
 A considerable theoretical work has shown how the electronic states near the Fermi level
 respond to the lattice and dopants organization changing the topology of the Fermi 
 surface \cite{tjapl,jb,lanio,jbmb,jbb,tj3,js,tj11}.
The interest is driven by the perspective that a 
the quasi one-dimensional ordering of dopants could generate stripes giving
a new incipient quasi 1D  electronic structure at the Fermi level which 
is expected  to enhance the critical temperature. 
The stripes periodicity driven by oxygen interstitial self organization sets up a potential
modulation, where the chemical potential is tuned at a Lifshitz transition for the appearing of a new 1D band.
The lattice structure of the stripes and the charge density have to
to be tuned so that one of the peaks of the density of states is near the position of the chemical potential.
Doping is crucial for high-$T_c$ superconductivity. 
It can be controlled either by the substitution of
the heavy atoms with different valency, like Sr (or Ba) for La in La$_{2-x}$Sr$_x$CuO$_{4}$.
or by varying the oxygen content 
which has been proven to be efficient for doping, either as a
vacancy or as an interstitial impurity.
Superconductivity can be enhanced by ordering of oxygen interstitials in cuprates like
La$_2$CuO$_{4 \pm \delta}$   \cite{poccia1}. 
 Band calculations show that oxygen vacancies in the apical positions in 
Ba$_2$CuO$_{4 - \delta}$ (BCO, with $\delta \approx 1$) make its electronic structure very similar to that of
 optimally doped La$_2$CuO$_{4}$ (LCO) \cite{jbmb,jbb}.
The $T_c$ of BCO is reported to be much larger than in LCO \cite{gao}. 
Self organization of oxygen interstitials enhances $T_c$ \cite{poccia1}, and the Fermi
surface (FS) can become fragmented by oxygen self organization \cite{jb}.
However, the exact role of $ordering$ of the defects is not well known in many cases.

 Recently experimental results have been reported on self organization of oxygen interstitials
 in doped cuprates HgBa$_2$CuO$_{4+y}$   \cite{campi3,campi4}. by scanning micro x-ray diffraction
which provide complementary information on local nanoscale structure investigation using x-ray absorption spectroscopy 
\cite{xaneshg1,xaneshg2,xaneshg3} using  EXAFS and XANES  methods \cite{doniach,garcia1}
which probe the deviation of the local structure from the average structure.
In this work we present electronic structure results for the hole-doped oxygen-enriched 
 HgBa$_2$CuO$_{4.167}$. Ordered O-stripes with 2 impurities along (0,1,0) are separated by 5 unit cells
 along (1,0,0). Cells with disordered distribution of the two O impurities and without impurities
 are studied for comparison.
The method of calculation is presented in sect. II.  Experimental information about the structure is used
to define the supercells of O-rich LNO, as is also discussed in sect. II.  
In sect. III we discuss the results of the calculations, and some ideas for future works are given together with the 
conclusions in sect. IV.

\section{Method of calculation.}

The calculations are made using the linear muffin-tin orbital (LMTO) method \cite{lmto,bdj} and the
local spin-density approximation (LSDA) \cite{lsda1,lsda2}. 
The details of the methods have been published earlier 
\cite{bj94,jb,tj1,tj11}.
The supercell Hg$_{12}$Ba$_{24}$Cu$_{12}$O$_{48+2}$ is extended 6- and 2-lattice constants
along $x$ and $y$, respectively. The 2 additional oxygens are inserted in the Hg-plane, as is
known to be the position of excess O in HBCO. These oxygen interstitials form a stripe running along $y$. 
Calculations for 
the elementary cell of HBCO need 8 atomic sites and 5 "empty spheres", which are included in 
 the most open part of the structure, see ref. \cite{bj94}. The empty sphere in the Hg plane,
 at (0.5, 0.5,0), is the 
location of excess oxygen. 
The lattice constant $a_0$=3.87 \AA, and c/a=2.445.
The elementary cell is extended 6$a_0$ along $x$ and 2$a_0$ along $y$. The empty spheres at
(0.5,0.5,0) and (0.5,1.5,0), and 
at (0.5,0.5,0) and (2.5,1.5,0) are occupied by O in the "stripe" supercell and
"disordered" supercells, respectively. This is in the latter case the most distant and uncorrelated
choice for the two interstitial O impurities. 
The band calculations are made for this supercell containing 156 sites totally, and
the basis set goes
up through $\ell$=2 for atoms and $\ell=1$ for empty spheres. 
Self-consistent, paramagnetic and spin-polarized calculations are made for these cells using
54 k-points. The FS plots used a finer k-point mesh, 210 k-points, in one plane half-way of
the maximum $k_z$.

Correlation is not expected to be an issue for cuprates and nickelates with doping well away from half-filling of the d-band. 
This is confirmed for cuprates from ARPES (angular-resolved photoemission
spectroscopy) and ACAR (angular correlation of positron annihilation radiation), which detect FS's and
bands that evolves with high doping in agreement with DFT (density-functional theory)
calculations \cite{pick,dama,posi}. 

\begin{table}[ht]
\caption{\label{tab1}
Local decomposition of the DOS at $E_F$ (in units of $(atom \cdot eV)^{-1}$) for 
Hg$_{12}$Ba$_{24}$Cu$_{12}$O$_{48}$ and "striped" and "disordered" Hg$_{12}$Ba$_{24}$Cu$_{12}$O$_{48+2}$.
The total $N(E_F)$ are 11.0, 37.0 and 47.5 $(cell \cdot eV)^{-1}$, respectively. Apical and planar O are 
indicated O$_{ap}$ and O$_{pl}$, respectively.
The additional impurity oxygens (O$_i$) have very large p-DOS, but also all atoms in the first layer
around an impurity get large DOS values.
  }
  \vskip 2mm
  \begin{center}
  \begin{tabular}{l c c c c c c c }
  \hline
  ~ & ~ & Ba & Hg & Cu & O$_{ap}$ & O$_{pl}$ & O$_i$  \\
  \hline \hline

 undoped & ~ & - & - & 0.73 & - & 0.12 &  ~  \\
 \hline

 stripe  & near O$_i$ & 0.02 & 0.84 & 1.0 & 0.50 & 0.21 & 5.5 \\
         & far from O$_i$ & - & - & 0.93 & 0.15 & 0.18 & ~ \\
 \hline
   disord.  & near O$_i$ & 0.3 & 0.46 & 1.1 & 0.29 & 0.20 & 8.0 \\
           & far from O$_i$ & - & - & 0.90 & 0.03 & 0.15 & ~ \\

  \hline
  \end{tabular}
  \end{center}
  \end{table}

\begin{figure}
\includegraphics[height=8.0cm,width=9.2cm]{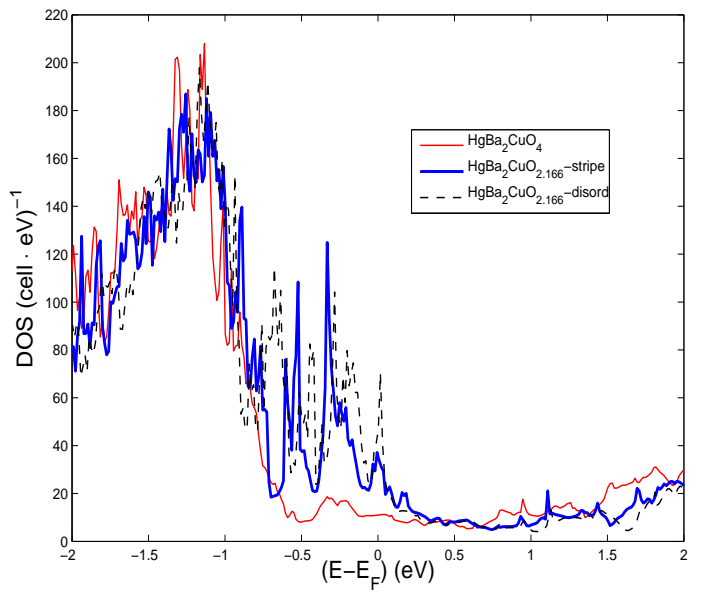}
\caption{(Color online)  The total DOS for Hg$_{12}$Ba$_24$Cu$_{12}$O$_{48+N}$ with $N$=0 (thin red) and $N$=2
(bold blue).
The (black) broken line is when the 2 O atoms occupy sites far from each other ("disordered"). }
\label{fig1}
\end{figure}

\begin{figure}
\includegraphics[height=8.0cm,width=9.2cm]{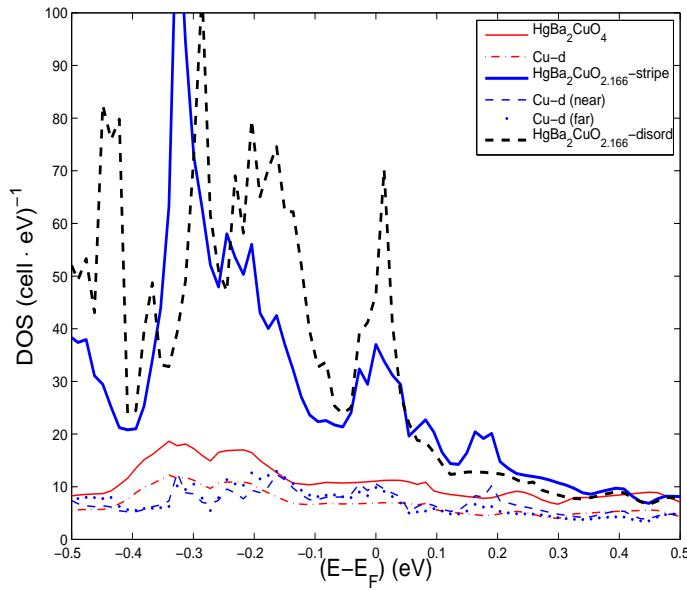}
\caption{(Color online)  The DOS functions in Fig. \ref{fig1} on a finer energy scale.
The partial Cu-d DOS are also shown for the undoped case (red semi-broken) and the striped case
for Cu near to (blue dotted) and far from (blue thin broken) the O$_i$-stripe. The partial Cu-d
DOS function are multiplied by 12 in order to be normalized to the total DOS. }
\label{fig1b}
\end{figure}

\section{Results and discussion.}

 The nonmagnetic (NM) total DOS functions at the Fermi level for the three supercells are 
shown  in Figs. \ref{fig1}. The undoped DOS agree with the DOS calculated previously
for one unit cell of HBCO using also $\ell=3$ states for the Ba sites \cite{bj94}. 
A difference is that the less dense
k-point mesh for the supercell makes the DOS curve less smooth. The total DOS at $E_F$ per elementary cell 
is about 1.0 $(eV)^{-1}$ compared to 0.92 here for the supercell.
A calculation of a supercell of intermediate size (Hg$_{4}$Ba$_{8}$Cu$_{4}$O$_{17}$)
corresponding to an impurity concentration of 0.25 shows that the DOS at $E_F$ increases by a factor
of two to about 1.8 per elementary cell \cite{js}. All atoms are close to the impurity in that case,
which explains that the local peaks in the states are not so narrow as in the present case. Here
$E_F$ is on a narrow peak in both cells with O impurities, which makes the DOS higher,
about 3 and 4 $(eV)^{-1}$ per elementary cell for the striped and disordered case respectively.
The increase of the DOS is limited to the first layers of atoms adjascent to the impurity, see
Table \ref{tab1}.
Hybridization with the p-states on the oxygen impurity atoms makes large s- and d-DOS on Hg, and
large increase of the p-DOS on apical oxygen states, while the changes on planar O and Ba are
not large. The Cu sites are quite distant to the O$_i$. Nevertheless,
the Cu d-DOS goes up by about 35 percent near
the impurity, and by 25 percent on the more distant Cu. 
This can be understood as an effect of hole doping from O$_i$ (see later) when $E_F$ is approaching a
van-Hove DOS peak as in other cuprates. More distant atoms of the other types
(Ba, Hg, and O) have their local DOS very much like those of the undoped HBCO,
i.e. only very small local DOS values. Already
the second layer from the O-stripe seems to have restored its local character as in undoped HBCO,
except for the modest increase of the Cu-d DOS.

The doping is important for the high-$T_c$ in HBCO, but it is not clear if this is due to the
high Cu-d DOS (and the presumed conservation of a FS cylinder, see later), or if the
very large increase of the DOS close to the impurity is the essential point. The grouping of the impurities
into stripes, instead of having them spread out as in the disorder case, permits to have
the long-range effect on Cu-d. In fact, the local Cu-d DOS is essentially unchanged among all Cu
in the striped case, it is 7 percent lower on the 3rd layer and 3.5 percent lower on the 2nd layer
compared to the 1st layer from the impurity. The variation of the local Cu-d DOS in the disordered case
is a bit larger, at most 20 percent (no sites go beyond the 2nd layer). From these results
one may suspect that the important mechanism of O-doping is to enforce the Cu-d DOS and
the generic cylindrical FS that is typical for the high $T_C$ cuprates \cite{js,pick}, and by grouping the impurities
into stripes or clusters one can avoid to have bad short-range effects from the impurity itself.
   
\begin{figure}
\includegraphics[height=8.0cm,width=9.0cm]{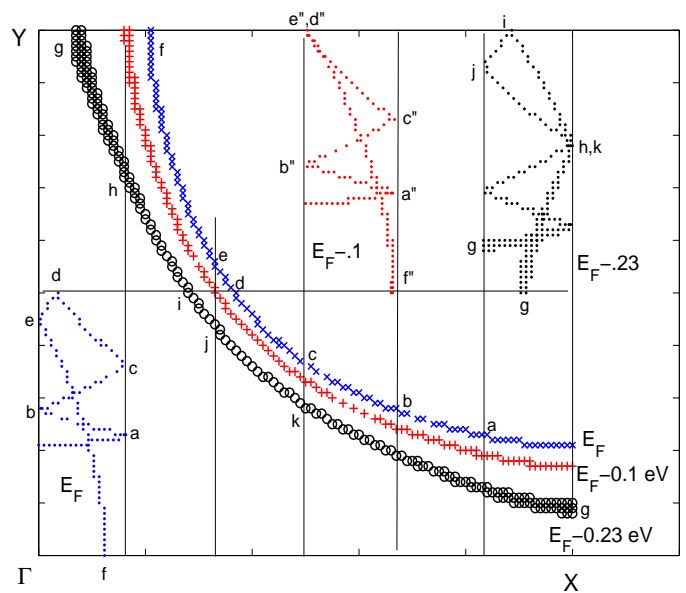}
\caption{(Color online) Upper large panel ($\Gamma$-$X$-$Y$) show the Fermi surface for one unitcell
of undoped HBCO. The "x" is at the calculated $E_F$, the "+" for 0.1 eV
down shifted, and "o" for 0.23 eV down shifted $E_F$. The small panels show how these FS look
after folding the bands into the BZ for the 6x2 supercell.
}
\label{fig2}
\end{figure}

\begin{figure}
\includegraphics[height=8.0cm,width=9.0cm]{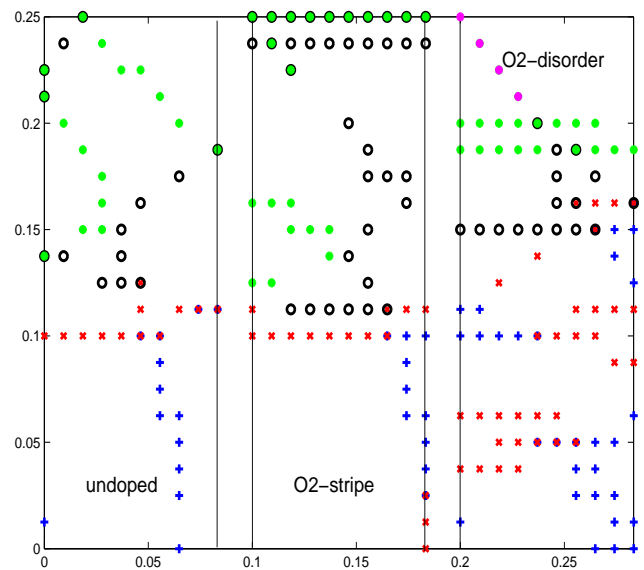}
\caption{(Color online) The FS of Hg$_{12}$Ba$_24$Cu$_{12}$O$_{48+N}$ for N=0 and 2 (as stripe and "disordered").
The different marks indicate different bands in the supercell band structure.
It is seen that the undoped FS agree well with what is expected, i.e the folded FS shown at the bottom left
of Fig. \ref{fig3}. The FS for the supercell with the O-stripe agree in part with the
folded, downshifted (0.1 eV) FS shown in Fig. \ref{fig3}, and hence the doping has induced
a hole doping. However, there is a new band and FS at the top that cannot be explained by from the
simple folding.
}
\label{fig3}
\end{figure}

 The changes in the FS from doping is studied by comparing the down-folded FS of the elementary
cell into the Brillouin Zone (BZ) for the supercell. This is easier to do than the instructive
method of folding supercell FS's into the BZ for the elementary cell \cite{sunko}.
The first step is to fold down the elementary FS of the large BZ into the smaller BZ
of the corresponding supercells. The result of this procedure is shown in Fig. \ref{fig2}
for 3 levels of $E_F$ corresponding to different hole doping. Fig. \ref{fig3} shows the 3 FS's
for the undoped, striped and disordered supercells in a plane halfway between the BZ center and
the maximum of $k_z$. The bands are calculated in 210 k-points, and the dots in the Fig. indicate
where a band energy is
within a small energy window around $E_F$.  The energy window is narrower in the high-DOS cases
in order to have not too many dots. The left panel shows the FS for the undoped supercell.
As expected, this FS is close to that shown in the lower left part of Fig \ref{fig2}. Practically
all features of the FS agree. The middle panel, showing the FS of the striped supercell,
can be compared with the down-folded middle panel in Fig. \ref{fig2}, which is labeled
"$E_F -.1$". The FS feature $f$ in the original FS, which show up at $f$ in the down-folded FS
without doping (lower left), has moved to $f"$ at this hole doping. Also other pieces of FS,
such as the merging of the points $e$ and $d$ into $e"$ and $d"$,
confirm the higher level of hole doping for the basic FS cylinder.
This is also corroborated by the effective valence charge within the Cu spheres, which
on the average decreases by 0.04 el./Cu, see Table \ref{tab2}. The Cu closest to O$_i$ has
a larger reduction, about 0.06 el/Cu.
However, on top of the FS panel for the striped supercell there is a horizontal local array of dots
that cannot be seen from the down-folded simple FS. A large DOS contribution is coming
from these points, and it is likely due to the O-i  band. The right panel of Fig. \ref{fig3}
shows the FS for the disordered supercell.  Here it is not easy to see resemblance
with any of the down-folded FS. The wide structure in the lower part of the panel, and the horizontal
lines in the middle, have no correspondences in Fig. \ref{fig2}. This suggests that when the O-doping
is high enough, and the O-i are spread out to allow for some electronic overlap between them,
they tend to destroy much of the original cuprate FS. The effective hole doping is slightly
larger than for the striped case, about 0.05 el./Cu, with the largest value still
at 0.06 el./Cu for the Cu closest to O$_i$. 

These results suggest that if the oxygen interstitials are well separated
they can still provide an essential hole doping for the rest of the lattice, while the local
perturbation near the interstitial ion will only have small effects on the region in between.
In this case there could be that superconductivity occurs in a filamentary region
which is separated from the stripes.

The calculated local DOS distributions in different regions 
suggest that it can be energetically favorable for the impurities to form clusters
or stripes rather than to have them isolated and spread out. The (hole) charge transfer ($\Delta Q$) to the interior
region far from the stripes,
$\Delta Q = N_i \cdot \Delta E$ leads to a gain of kinetic energy from that region;
$\Delta E_i=\Delta Q \cdot \Delta\epsilon_i$.
The local DOS $N_i$ is small here, which makes $\Delta\epsilon_i$ and $\Delta E_i$ large.
Since the DOS at the stripe, $N_s$, is large, there will be a small loss in kinetic energy ($\Delta E_s$) from this region;
$\Delta E_s = \Delta Q \cdot \Delta\epsilon_s$, since $\Delta\epsilon_s$ is small, and totally there 
is a kinetic energy gain.
However, the DOS will be equally high everywhere if the distribution of O-i is uniform, and the gain and loss
of kinetic energy will cancel even if there is a transfer between different regions. 
These arguments are just indicative, since
hybridization and potential energy contributions are missing.

  \begin{table}[ht]
\caption{\label{tab2}
Absolute values of local moments ($\mu_B/Cu$) on Cu in spin polarized calculations where magnetic fields $\pm$ 0.4 $eV$
are applied on all Cu in an AFM pattern. The last column shows Q$_{Cu}$, the average
number of valence electrons with the Cu MT-sphere. }
  \vskip 2mm
  \begin{center}
  \begin{tabular}{l c c c c}
  \hline
 ~  & average & near O$_i$ & far from O$_i$ & Q$_{Cu}$ \\
  \hline \hline

undoped    & 0.30 & -  & - & 10.40 \\
striped    & 0.29 & 0.29  & 0.30 & 10.36 \\
disordered & 0.29 & 0.28  & 0.30 & 10.35 \\

  \hline
  \end{tabular}
  \end{center}
  \end{table}

 The large difference in the DOS between the 3 cases suggests that spin-fluctuations
 could appear more easily in the impurity phases. This should be the case for ferromagnetism
 (FM) according to the Stoner model. However, anti-ferromagnetism (AFM) is stabilized in undoped
cuprates and AFM spin-fluctuations are likely in hole doped cuprate systems 
especially when they are coupled to phonon distortions \cite{tj7,tj6}. Thermal disorder
can alter local magnetic moments and contribute to spin fluctuations \cite{eri14,cevib}. 

It is not certain that a high DOS will promote AFM fluctuations, and in order to compare the
ability for AFM spin fluctuations in the 3 cases we perform
spin-polarized calculations where identical AFM fields are applied on all Cu. 
The results show very small differences in the AFM response between undoped, striped and
disordered supercells. The reason is probably that the electronic overlap between O-i and Cu is
small since they are in different layers.

 The average absolute value of the local moment on Cu is highest
in the undoped case, see Table \ref{tab2}. This is consistent with the fact that real undoped
cuprates are AFM insulators, while with hole doping the systems become metallic, and the
stable AFM is gradually replaced by AFM fluctuations. 

\section{Conclusion.}

Stripes of O-i dopants in mercury based cuprates should contribute 
to a potential modulation with the formation of a new quasi 1D band with a peak formation in the DOS at the band edge. 
The spatial regions far from the O-i wires and the immediate impurity clusters are spatially and electronically separated.
The very high DOS within the stripes indicate that its local electronic structure is 
very different from the rest of the cuprate. It is temping to view the stripe as a
highly conducting "wire" within a typical cuprate which is less conducting.
It is possible that high temperature superconductivity takes place in a $third$ intermediate filamentary region intermediate 
the $first$ O-i rich wires, highly doped regions, formed by clustering of O interstitials and the $second$ spacial regions oxygen-poor empty of the dopant wires where spin fluctuations are expected.
This scenario should be verified or discarded by space resolved angular resolved photoemission ARPES experiments, tunneling or optical spectroscopy
on well defined surfaces using nanobeams.

\end{document}